\documentclass[onecolumn,amsmath,amssymb]{qm2008_abs}
\usepackage{graphicx}
\usepackage{dcolumn}
\usepackage{bm}
\begin{document}

\title{{\Large 
System size, energy, centrality and pseudorapidity
dependence of charged-particle density\\
in Au+Au and Cu+Cu collisions at RHIC}}

\bigskip
\bigskip
\author{\large G\'abor I. Veres (for the PHOBOS Collaboration)}
\email{vg@ludens.elte.hu}
\affiliation{E\"otv\"os Lor\'and University, Budapest, H-1111, Hungary\\
Massachusetts Institute of Technology, Cambridge, MA 02139, USA}
%
%
\author{B.Alver}\affiliation{Massachusetts Institute of Technology, Cambridge, MA 02139-4307, USA}
\author{B.B.Back}\affiliation{Argonne National Laboratory, Argonne, IL 60439-4843, USA}
\author{M.D.Baker}\affiliation{Brookhaven National Laboratory, Upton, NY 11973-5000, USA}
\author{M.Ballintijn}\affiliation{Massachusetts Institute of Technology, Cambridge, MA 02139-4307, USA}
\author{D.S.Barton}\affiliation{Brookhaven National Laboratory, Upton, NY 11973-5000, USA}
\author{R.R.Betts}\affiliation{University of Illinois at Chicago, Chicago, IL 60607-7059, USA}
\author{A.A.Bickley}\affiliation{University of Maryland, College Park, MD 20742, USA}
\author{R.Bindel}\affiliation{University of Maryland, College Park, MD 20742, USA}
\author{W.Busza}\affiliation{Massachusetts Institute of Technology, Cambridge, MA 02139-4307, USA}
\author{A.Carroll}\affiliation{Brookhaven National Laboratory, Upton, NY 11973-5000, USA}
\author{Z.Chai}\affiliation{Brookhaven National Laboratory, Upton, NY 11973-5000, USA}
\author{V.Chetluru}\affiliation{University of Illinois at Chicago, Chicago, IL 60607-7059, USA}
\author{M.P.Decowski}\affiliation{Massachusetts Institute of Technology, Cambridge, MA 02139-4307, USA}
\author{E.Garc\'{\i}a}\affiliation{University of Illinois at Chicago, Chicago, IL 60607-7059, USA}
\author{T.Gburek}\affiliation{Institute of Nuclear Physics PAN, Krak\'{o}w, Poland}
\author{N.George}\affiliation{Brookhaven National Laboratory, Upton, NY 11973-5000, USA}
\author{K.Gulbrandsen}\affiliation{Massachusetts Institute of Technology, Cambridge, MA 02139-4307, USA}
\author{C.Halliwell}\affiliation{University of Illinois at Chicago, Chicago, IL 60607-7059, USA}
\author{J.Hamblen}\affiliation{University of Rochester, Rochester, NY 14627, USA}
\author{I.Harnarine}\affiliation{University of Illinois at Chicago, Chicago, IL 60607-7059, USA}
\author{M.Hauer}\affiliation{Brookhaven National Laboratory, Upton, NY 11973-5000, USA}
\author{C.Henderson}\affiliation{Massachusetts Institute of Technology, Cambridge, MA 02139-4307, USA}
\author{D.J.Hofman}\affiliation{University of Illinois at Chicago, Chicago, IL 60607-7059, USA}
\author{R.S.Hollis}\affiliation{University of Illinois at Chicago, Chicago, IL 60607-7059, USA}
\author{R.Ho\l y\'{n}ski}\affiliation{Institute of Nuclear Physics PAN, Krak\'{o}w, Poland}
\author{B.Holzman}\affiliation{Brookhaven National Laboratory, Upton, NY 11973-5000, USA}
\author{A.Iordanova}\affiliation{University of Illinois at Chicago, Chicago, IL 60607-7059, USA}
\author{E.Johnson}\affiliation{University of Rochester, Rochester, NY 14627, USA}
\author{J.L.Kane}\affiliation{Massachusetts Institute of Technology, Cambridge, MA 02139-4307, USA}
\author{N.Khan}\affiliation{University of Rochester, Rochester, NY 14627, USA}
\author{P.Kulinich}\affiliation{Massachusetts Institute of Technology, Cambridge, MA 02139-4307, USA}
\author{C.M.Kuo}\affiliation{National Central University, Chung-Li, Taiwan}
\author{W.Li}\affiliation{Massachusetts Institute of Technology, Cambridge, MA 02139-4307, USA}
\author{W.T.Lin}\affiliation{National Central University, Chung-Li, Taiwan}
\author{C.Loizides}\affiliation{Massachusetts Institute of Technology, Cambridge, MA 02139-4307, USA}
\author{S.Manly}\affiliation{University of Rochester, Rochester, NY 14627, USA}
\author{A.C.Mignerey}\affiliation{University of Maryland, College Park, MD 20742, USA}
\author{R.Nouicer}\affiliation{Brookhaven National Laboratory, Upton, NY 11973-5000, USA}
\author{A.Olszewski}\affiliation{Institute of Nuclear Physics PAN, Krak\'{o}w, Poland}
\author{R.Pak}\affiliation{Brookhaven National Laboratory, Upton, NY 11973-5000, USA}
\author{C.Reed}\affiliation{Massachusetts Institute of Technology, Cambridge, MA 02139-4307, USA}
\author{E.Richardson}\affiliation{University of Maryland, College Park, MD 20742, USA}
\author{C.Roland}\affiliation{Massachusetts Institute of Technology, Cambridge, MA 02139-4307, USA}
\author{G.Roland}\affiliation{Massachusetts Institute of Technology, Cambridge, MA 02139-4307, USA}
\author{J.Sagerer}\affiliation{University of Illinois at Chicago, Chicago, IL 60607-7059, USA}
\author{H.Seals}\affiliation{Brookhaven National Laboratory, Upton, NY 11973-5000, USA}
\author{I.Sedykh}\affiliation{Brookhaven National Laboratory, Upton, NY 11973-5000, USA}
\author{C.E.Smith}\affiliation{University of Illinois at Chicago, Chicago, IL 60607-7059, USA}
\author{M.A.Stankiewicz}\affiliation{Brookhaven National Laboratory, Upton, NY 11973-5000, USA}
\author{P.Steinberg}\affiliation{Brookhaven National Laboratory, Upton, NY 11973-5000, USA}
\author{G.S.F.Stephans}\affiliation{Massachusetts Institute of Technology, Cambridge, MA 02139-4307, USA}
\author{A.Sukhanov}\affiliation{Brookhaven National Laboratory, Upton, NY 11973-5000, USA}
\author{A.Szostak}\affiliation{Brookhaven National Laboratory, Upton, NY 11973-5000, USA}
\author{M.B.Tonjes}\affiliation{University of Maryland, College Park, MD 20742, USA}
\author{A.Trzupek}\affiliation{Institute of Nuclear Physics PAN, Krak\'{o}w, Poland}
\author{C.Vale}\affiliation{Massachusetts Institute of Technology, Cambridge, MA 02139-4307, USA}
\author{G.J.van~Nieuwenhuizen}\affiliation{Massachusetts Institute of Technology, Cambridge, MA 02139-4307, USA}
\author{S.S.Vaurynovich}\affiliation{Massachusetts Institute of Technology, Cambridge, MA 02139-4307, USA}
\author{R.Verdier}\affiliation{Massachusetts Institute of Technology, Cambridge, MA 02139-4307, USA}
\author{G.I.Veres}\affiliation{Massachusetts Institute of Technology, Cambridge, MA 02139-4307, USA}
\author{P.Walters}\affiliation{University of Rochester, Rochester, NY 14627, USA}
\author{E.Wenger}\affiliation{Massachusetts Institute of Technology, Cambridge, MA 02139-4307, USA}
\author{D.Willhelm}\affiliation{University of Maryland, College Park, MD 20742, USA}
\author{F.L.H.Wolfs}\affiliation{University of Rochester, Rochester, NY 14627, USA}
\author{B.Wosiek}\affiliation{Institute of Nuclear Physics PAN, Krak\'{o}w, Poland}
\author{K.Wo\'{z}niak}\affiliation{Institute of Nuclear Physics PAN, Krak\'{o}w, Poland}
\author{S.Wyngaardt}\affiliation{Brookhaven National Laboratory, Upton, NY 11973-5000, USA}
\author{B.Wys\l ouch}\affiliation{Massachusetts Institute of Technology, Cambridge, MA 02139-4307, USA}

\bigskip
\bigskip

\begin{abstract}
\leftskip1.0cm
\rightskip1.0cm
Charged particle pseudorapidity distributions are presented
from the PHOBOS experiment at RHIC, measured in Au+Au and Cu+Cu 
collisions at $\sqrt{s_{_{\rm NN}}}$=19.6, 22.4, 62.4, 130 and 200~GeV, 
as a function of collision centrality. The presentation includes the 
recently analyzed Cu+Cu data at 22.4 GeV. The 
measurements were made 
by the same detector setup over a broad range in pseudorapidity, 
$|\eta|<5.4$, allowing for a reliable systematic study of particle 
production as a function of energy, centrality and system size. 
Comparing Cu+Cu and Au+Au results, we find that the total number of 
produced charged particles and the overall shape (height and width) of 
the pseudorapidity distributions are determined by the number of nucleon 
participants, $N_{part}$. Detailed
comparisons reveal that the matching of the shape of the Cu+Cu and Au+Au
pseudorapidity distributions over the full range of $\eta$ is better
for the same $N_{part}/2A$ value than for the same $N_{part}$ 
value, where $A$ denotes the mass number. 
In other words, it is the geometry of the nuclear overlap zone, rather 
than just the number of nucleon participants that drives the detailed 
shape of the pseudorapidity distribution and its centrality dependence.
\end{abstract}
\maketitle

\section{Introduction}

The charged particle pseudorapidity distribution, $dN_{ch}/d\eta$, is a 
well defined experimental quantity that reflects the initial conditions 
of the system, e.g. parton shadowing and gluon saturation, and also the 
effects of rescattering and hadronic final
state interactions: it represents the time-integral of the particle
production throughout the entire collision. With the advent of Cu+Cu 
collisions at RHIC, the system size dependence of important observables
can be studied using different collision geometries. The Cu+Cu
results \cite{phobos1} test the simple scaling features observed previously in
Au+Au collisions \cite{brahms2,phobos3}. They significantly extend 
the $N_{part}$ range measured in Au+Au collisions, while the two 
systems can also be compared at the same $N_{part}$, as illustrated below.

\section{Experimental setup and data analysis}

The Cu+Cu data were collected with the multiplicity array of the
PHOBOS detector \cite{phobos5} during the RHIC 2005 run. The array 
consists of single-layered silicon sensors assembled into a long, 
tube-shaped Octagon detector surrounding the collision point, and into 
three Ring sensors on each side, detecting large-$|\eta|$ particles. 
Simulations of the detector performance were based on the HIJING event 
generator and GEANT, including the response of the scintillator Paddle 
trigger counters.

Data from the Cu+Cu and Au+Au collisions were analyzed using the 
`hit-counting' and `analog' methods \cite{phobos6}. The latter was
corrected for multiply-charged fragments emitted at large $\eta$. This
correction decreases with centrality and collision energy, and it is 
less than 6\% of the total number of charged particles.

The estimated trigger efficiency is 
84$\pm$5\% and 75$\pm$5\% in Cu+Cu collisions at 200 and 62.4~GeV. 
The centrality of the collision was estimated from the Paddle 
scintillator signals. 
At 22.4 and 19.6~GeV, the pathlength-corrected energy sum 
\cite{phobos3} deposited in the Octagon was used ($|\eta|<3.2$). 
A Glauber-model calculation was employed to estimate 
$\langle N_{part} \rangle$ for each centrality bin.

\section{Results}

\begin{figure}
\begin{center}
\includegraphics[width=50mm]{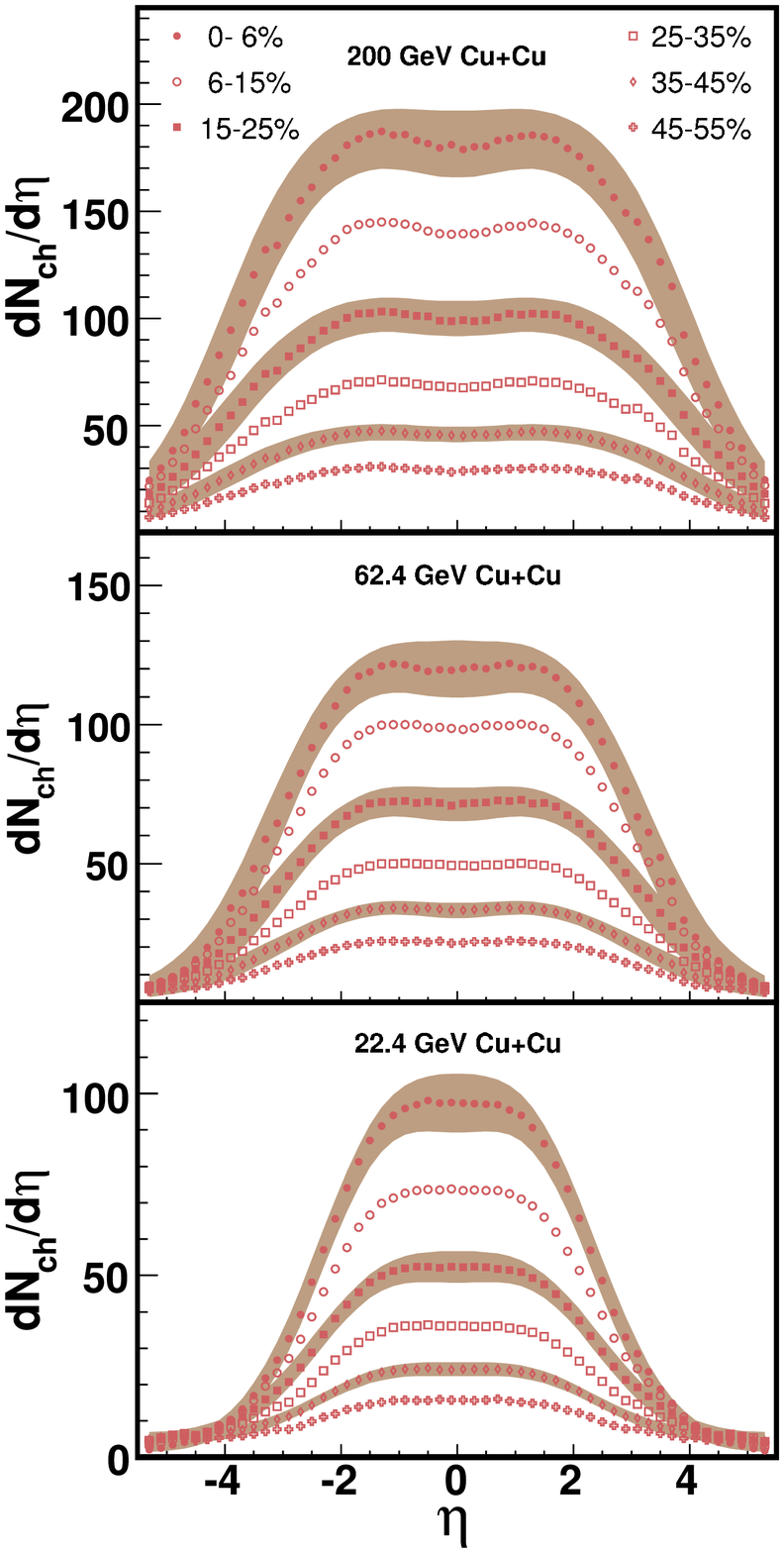}
\includegraphics[width=50mm]{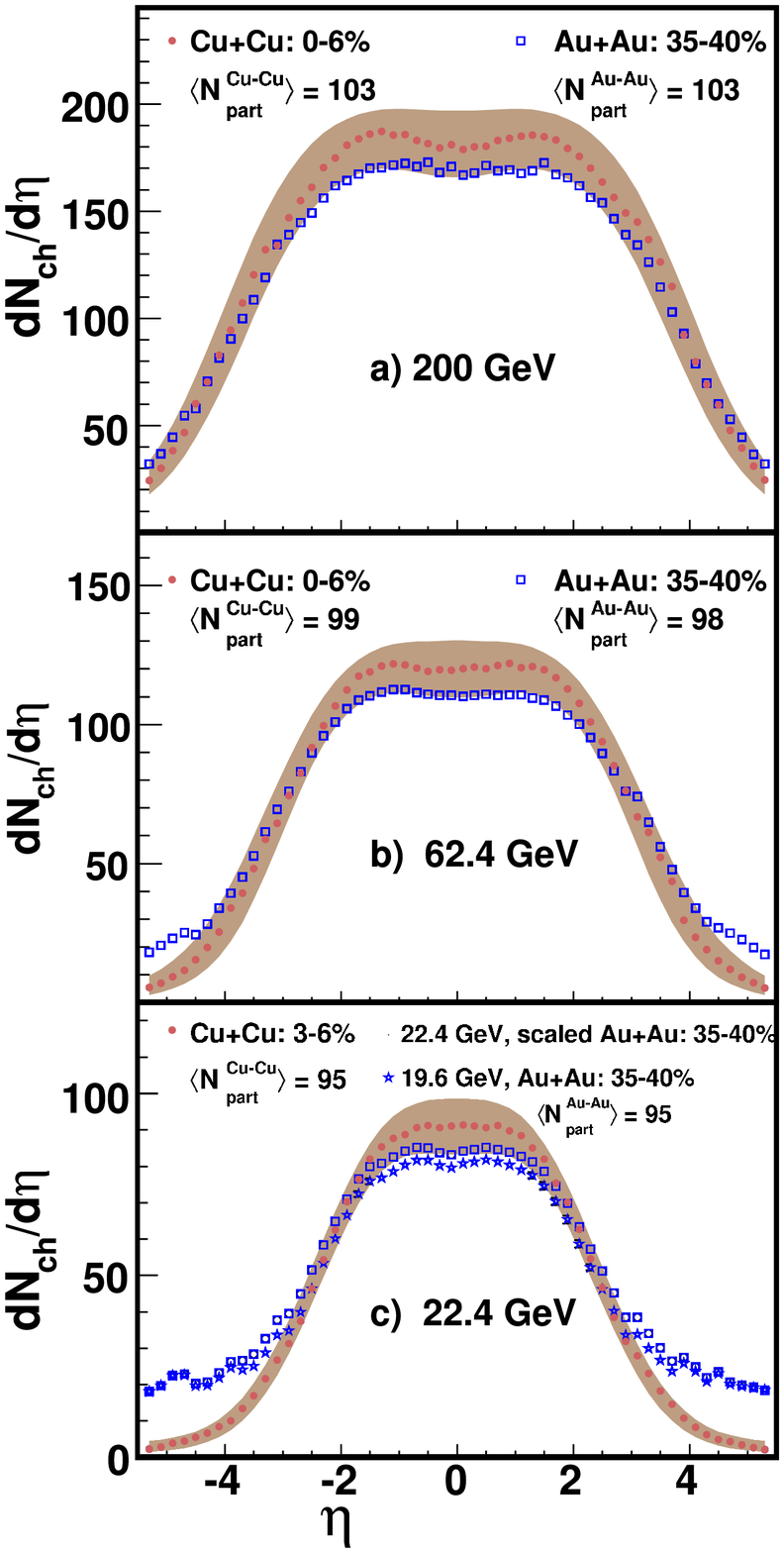}
\caption{Left panel: pseudorapidity distributions of primary charged 
particles from Cu+Cu collisions at 22.4, 62.4 and 200~GeV collision 
energy per nucleon pair for various centrality bins. Right panel: 
comparison of $dN_{ch}/d\eta$ distributions in Cu+Cu and Au+Au 
collisions - corresponding to the same $\langle N_{part}\rangle$. 
90\% C.L. systematic errors are shown as bands.}
\label{fig1}
\end{center}
\end{figure}

The $dN_{ch}/d\eta$ distributions in Cu+Cu collisions for various
collision energies and centralities are shown in Fig. \ref{fig1}. On the 
right panel the Cu+Cu and Au+Au collisions are compared, where the 
centrality bins are chosen such that $\langle N_{part} \rangle$ in both 
systems are similar. One can conclude that although the distributions 
agree at the same $\langle N_{part} \rangle$ to first order, there are 
differences at large $|\eta|$ and low energies. 
In this context, we note that the two nuclear spectator remnants are 
larger in Au+Au than in Cu+Cu collisions and thus may be playing a role.

\begin{figure}
\begin{center}
\includegraphics[width=100mm]{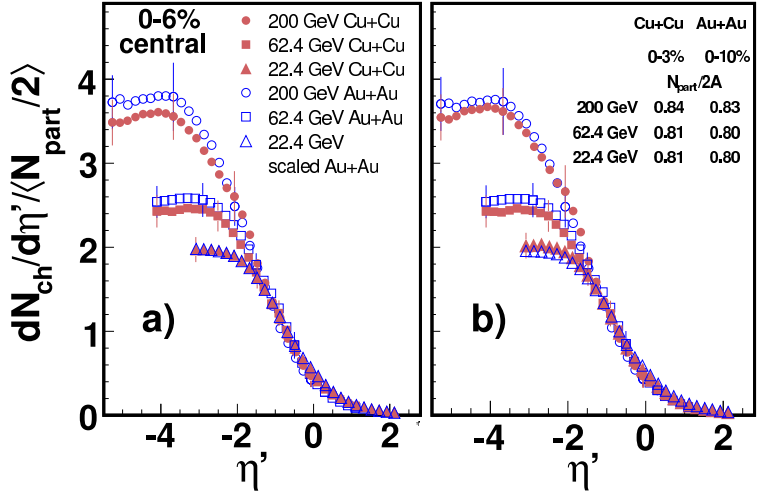}
\caption{Pseudorapidity distributions in Cu+Cu and Au+Au collisions at
various RHIC energies, normalized by the number of participant pairs,
plotted as a function of $\eta'=\eta-y_{beam}$, for a) the 6\% most 
central events and b) central events with similar $N_{part}/2A$. 90\% 
C.L. systematic errors are shown for a few typical data points.}
\label{fig2}
\end{center}
\end{figure}

The $dN_{ch}/d\eta$ distributions exhibit longitudinal scaling when 
observed from the rest frame of one of the colliding nuclei. The 
coordinate transformation to the `target' frame approximately 
corresponds to a shift by the beam rapidity, $y_{beam}$.
Figure \ref{fig2} compares the $dN_{ch}/d\eta'$ distributions (where 
$\eta'=\eta-y_{beam}$) after normalization by $N_{part}$: a) data from 
the Cu+Cu and Au+Au systems plotted at the same fraction of the total cross 
section (0-6\% most central bin), and b) at the same value of $N_{part}/2A$ 
(where $A$ is the mass number). Both cases indicate that the scaled 
particle density only depends on the collision energy 
and geometry, but not on the size of the nuclei.
The $dN_{ch}/d\eta'/\langle N_{part}\rangle$ distributions 
for the same centrality in both systems agree within errors, and the 
overall agreement improves if the centralities are compared on the 
basis of the $N_{part}/2A$ quantity (the fraction of participating 
nucleons).
The longitudinal scaling is similarly present in the Cu+Cu and in the
Au+Au data.

\begin{figure}
\begin{center}
\includegraphics[width=135mm]{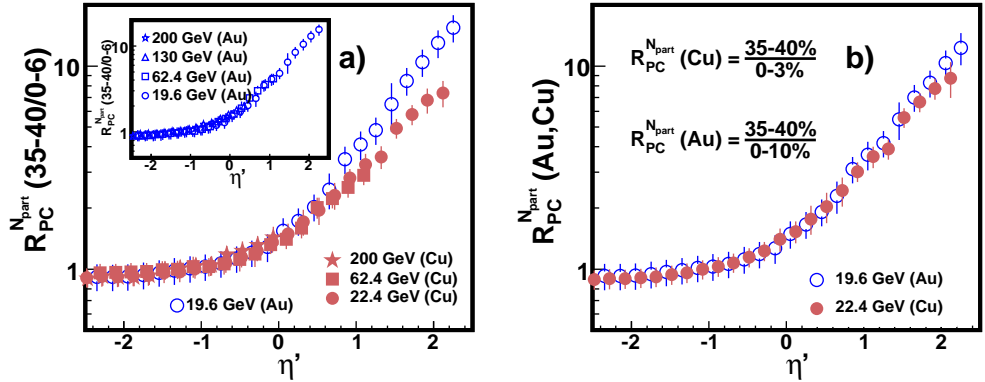}
\caption{The semi-peripheral to central 
$dN_{ch}/d\eta'/\langle N_{part}/2\rangle$ ratio for
Cu+Cu and Au+Au collisions at RHIC energies. The centrality bins are
selected a) according to fractional cross section (35-40\%/0-6\%) and
b) such that $N_{part}/2A$ is matched between the two systems. Inset: 
the same quantity plotted for Au+Au data for four different energies. 
The error bars represent 90\% C.L. systematic errors on the ratio.}
\label{fig3}
\end{center}
\end{figure}

The factorization between collision energy and centrality can be most
precisely studied by examining the ratios of the 
$dN/d\eta'/\langle N_{part}\rangle$ distributions in central and 
semi-central collisions, denoted by $R_{PC}^{N_{part}}$, at various 
energies. The 
published Au+Au results \cite{phobos4} are shown by the inset of
Fig.~\ref{fig3}a, exhibiting the same factorization feature as the 
recent Cu+Cu data. The above ratio for Cu+Cu data is similar to that in 
Au+Au data, except at the highest $\eta'$ values.
Fig.~\ref{fig3}b shows the $R_{PC}$ ratio for Cu+Cu and Au+Au for 
centrality bins where the $N_{part}/2A$ values are matched. The latter 
quantity characterizes the initial geometry more precisely, and indeed, 
the centrality evolution of the $dN_{ch}/d\eta$ distributions measured 
in Cu+Cu and Au+Au collisions are most similar if the centrality is 
quantified by $N_{part}/2A$.

\begin{figure}
\begin{center}
\includegraphics[width=71mm]{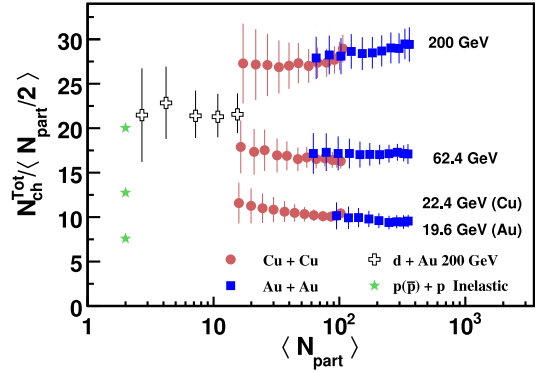}
\caption{
The integrated number of charged particles, scaled by $N_{part}/2$, in 
p+p, d+Au, Cu+Cu and Au+Au collisions as a function of centrality 
\cite{phobos4,phobos7}. The uncertainty of $N_{part}$ has been included 
in the error bars.}
\label{fig4}
\end{center}
\end{figure}

The total number of charged particles, $N_{tot}$, normalized
by $N_{part}$, is presented in Fig.~\ref{fig4} for Cu+Cu collisions at 
22.4, 62.4 and 200~GeV collision energy per nucleon pair, and compared 
to smaller (p+p, d+Au) and larger (Au+Au) systems as a function of 
centrality. One can conclude that $N_{tot}$ scales approximately 
linearly with $N_{part}$, and the normalized yield has similar values 
for the two heavy colliding systems. The d+Au data do not seem to 
interpolate smoothly between the p+p and heavy ion data points.

\section{Summary}

Charged particle $\eta$ distributions were presented,  
including the recently analysed Cu+Cu 
data taken at various collision energies.
The $dN/d\eta'$ distributions scaled by $N_{part}/2$, as well as their 
peripheral to central ratio were found to be
independent of the mass number, $A$, of the colliding nuclei if 
centrality classes with the same $N_{part}/2A$ (fraction of participant 
nucleons) are compared.

\noindent
{\bf Acknowledgements:} This work was partially supported by U.S. DOE grants
DE-AC02-98CH10886, DE-FG02-93ER40802, DE-FG02-94ER40818,
DE-FG02-94ER40865, DE-FG02-99ER41099, and
DE-AC02-06CH11357, by U.S. NSF grants 9603486, 0072204 and 0245011, by
Polish MNiSW grant N N02 282234 (2008-2010), by NSC of Taiwan Contract NSC
89-2112-M-008-024, by Hungarian grants OTKA F49823, NKTH-OTKA H07-C 
74248 and by the Magyary Postdoctoral Fellowship.


\end{document}